\documentclass[twoside]{article}
\usepackage{PRIMEarxiv}
\usepackage[utf8]{inputenc} 
\usepackage[T1]{fontenc}    
\usepackage{hyperref}       
\usepackage{url}            
\usepackage{booktabs}       
\usepackage{amsfonts}       
\usepackage{nicefrac}       
\usepackage{microtype}      
\usepackage{lipsum}
\usepackage{listings}
\usepackage{fancyhdr}       
\usepackage{graphicx}       
\graphicspath{{media/}}     
\usepackage{subcaption}
\usepackage{calc}
\usepackage{amssymb}
\usepackage{amstext}
\usepackage{amsmath}
\usepackage{amsthm}
\usepackage{multicol}
\usepackage{pslatex}
\usepackage{apalike}
\usepackage{algorithm2e}
\usepackage[bottom]{footmisc}
\usepackage{multirow}
\usepackage{xcolor}
\usepackage{verbatim}

\usepackage{color}
\usepackage[dvipsnames]{xcolor}
\usepackage{multirow}
\usepackage{amsmath} 
\usepackage{pgfplots}
\pgfplotsset{compat=1.18} 
\usepackage{tikz}

\begin{document}

\fancyhead[LO]{Financial News Summarization}
\fancyhead[RE]{Reche et al.} 

\title{Financial News Summarization: Can extractive methods still offer a true alternative to LLMs?}

\author{Nicolas Reche$^1$, Elvys Linhares-Pontes$^2$, Juan-Manuel Torres-Moreno$^1$\\
$^1$Laboratoire Informatique d'Avignon, AU\\Avignon, France\\
\texttt{nicolas.reche@alumni.univ-avignon.fr, juan-manuel.torres@univ-avignon.fr}\\
$^2$Trading Central Labs,  France
\\ \texttt{elvys.linharespontes@tradingcentral.com}
}

\maketitle              
\begin{abstract}
Financial markets change rapidly due to news, economic shifts, and geopolitical events. Quick reactions are vital for investors to avoid losses or capture short-term gains. As a result, concise financial news summaries are critical for decision-making. With over 50,000 financial articles published daily, automation in summarization is necessary. This study evaluates a range of summarization methods, from simple extractive techniques to advanced large language models (LLMs), using the FinLLMs Challenge dataset. LLMs generated more coherent and informative summaries, but they are resource-intensive and prone to hallucinations, which can introduce significant errors into financial summaries. In contrast, extractive methods perform well on short, well-structured texts and offer a more efficient alternative for this type of article. The best ROUGE results come from fine-tuned LLM model like FT-Mistral-7B, although our data corpus has limited reliability, which calls for cautious interpretation.
\end{abstract}


\section{Introduction}
\label{sec:intro}

Financial markets operate in an extremely dynamic and fast-paced environment, where conditions can change within minutes. Prices react almost instantly to news, economic data, or geopolitical events, forcing professionals, investors, and analysts to be highly responsive to avoid losses or seize fleeting opportunities. This rapid context is accompanied by an explosion in information volume, with a very large number of financial articles published daily by both specialized and general news sources. To manage this massive influx, financial monitoring platforms used by analysts, fund managers, and research teams must continuously process the overwhelming flow of articles, reports, press releases, and economic analyses. An ultra-concise summary format enables rapid scanning of vast amounts of information, helping to identify those that warrant deeper analysis and thus saving valuable time.

Financial documents vary widely: earnings calls, annual reports, press releases, and news articles, which complicates the application of universal summarization methods due to differences in length, structure, and information density. Studies have focused on specific formats such as call transcripts \cite{978-3-031-28238-6_1} or financial reports, highlighting the need for tailored approaches. Large Language Models (LLMs) excel at analysis and report generation \cite{wu2023bloomberggptlargelanguagemodel,li2024largelanguagemodelsfinance}, but they are costly, sometimes inaccurate, and require extensive fine-tuning. For instance, summarizing 50,000 articles at 3 seconds each would take over 41 hours, too slow for fast decision-making. In contrast, extractive methods, which are much faster and more economical, struggle to produce coherent summaries for long documents such as annual reports (Form 10-K).\footnote{Example of Form 10-K: \url{https://www.sec.gov/Archives/edgar/data/789019/ 000095017024087843/msft-20240630.htm}} However, for shorter articles from sources like Reuters or Bloomberg, they can remain a viable and scalable solution.

In this study, we systematically compare different summarization techniques, ranging from simple extractive methods to state-of-the-art LLMs, in order to evaluate their effectiveness at producing single-sentence summaries of financial news articles. Our objective is to determine whether the use of computationally expensive LLM-based approaches is justified for short financial documents. Alternatively, we explore if  lightweight extractive methods can generate concise and sufficiently informative summaries while significantly reducing computational cost.


Our study shows that for short financial news articles, simple extractive methods such as Lead-1 remain highly effective, partly due to the typical structure of these texts. While LLM-based models generate more fluent summaries, they require adaptation to the financial domain to achieve strong performance. Fine-tuning on domain-specific data significantly improves their results, highlighting the importance of context adaptation for automatic summarization tasks.

This article is organized as follows: 
we review related work on summarization techniques in Section~\ref{sec:related_work}. Section~\ref{sec:case_study} presents our case study, including the compared approaches. 
We report our experimental results in Section~\ref{sec:results}, followed by conclusions and future directions in Section~\ref{sec:CONC}.

\section{Related work}
\label{sec:related_work}

Automatic text summarization is a long-standing challenge in Natural Language Processing (NLP), with applications ranging from the summarization of general news articles to summarization tasks in specialized domains such as financial documents. In the latter, the rigid structure of documents, the density of information, and the importance of numerical precision introduce unique constraints for summarization.\vspace{2pt}

Two main approaches have historically structured the field: extractive summarization, which selects sentences directly from the source text, and abstractive summarization, which generates new formulations that synthesize the essential content ~\cite{linhares-pontes-etal-2018-multi}. However, in recent years, the boundaries between the two approaches have gradually blurred, particularly with the emergence of hybrid or contextually adaptive strategies.
Recent advancements in financial summarization have focused on the synthesis of long financial documents, such as annual reports and earnings call transcripts.\cite{el-haj-etal-2020-financial} organized the Financial Narrative Summarization (FNS) shared task, evaluating both paradigms (extractive and abstractive) on UK annual reports. \cite{abdaljalil-bouamor-2021-exploration} further explored summarization techniques tailored for financial reports, employing extractive summarization, NLP, and both machine and deep learning to generate structured summaries. These studies show that while extractive methods are robust, they struggle to produce readable and concise summaries when faced with highly verbose content. In this vein, the release of domain-specific datasets such as ECTSum \cite{mukherjee-etal-2022-ectsum} has accelerated research toward more structured summarization formats, such as bullet-point summaries, which are better aligned with professional use cases. 

More recent work focused on earnings call transcripts \cite{978-3-031-28238-6_1} reveals a methodological shift toward hybrid pipelines, combining unsupervised extractive modules with abstractive reformulations. The FLANFinBPS model exemplifies this dual-layer strategy, acknowledging the limits of unguided generation while leveraging the linguistic flexibility of generative models.

On the generative front, models such as PEGASUS ~\cite{pmlr-v119-zhang20ae} and Transformer-BiLSTM architectures with graph-based decoders \cite{10214663} have improved syntactic fluency and contextual awareness. However, their semantic reliability remains inconsistent when applied to financial texts without domain-specific adaptation.
This limitation has spurred the emergence of finance-specialized language models like BloombergGPT \cite{wu2023bloomberggptlargelanguagemodel} and FinGPT \cite{yang2023fingptopensourcefinanciallarge}. These models rely on massive proprietary corpora and demonstrate strong performance, yet they also raise concerns around accessibility, transparency, and scientific reproducibility. FinGPT, for instance, promotes an open-source, data-centric approach to address this issue.

In the broader context, general-purpose LLMs such as Mistral, DeepSeek, LLaMA, and OpenAI's GPT models have also played a crucial role in advancing NLP capabilities, including financial text understanding and summarization. These models have been further fine-tuned for financial-domain tasks such as summarization~\cite{pontes-etal-2024-l3itc}, text analysis~\cite{978-3-031-78255-8_1}, and other specialized applications, enhancing their effectiveness in financial NLP.

Recent methods such as GraphRAG \cite{shukla-etal-2025-graphrag} depart from sequential models by integrating knowledge graphs into retrieval-augmented generation (RAG) architectures. Their study highlighted the challenges of summarizing lengthy financial documents and proposed entity and relation extraction optimization to improve summarization performance.Although promising, these techniques remain underexplored in domains involving short, well-structured, and regularly published content, such as financial news articles.

\section{Study case}
\label{sec:case_study}

In this section we compare the performance of extractive approaches and distilled versions of LLMs in the task of summarizing short financial news documents. Our objective is to analyze the differences between these methodologies and assess the effectiveness of fine-tuning a general-purpose LLM for financial summarization.

This study also aims to explore the possibility of redefining the boundary between algorithmic sophistication and pragmatic efficiency in the financial domain, particularly in scenarios where summary quality must be balanced against computational cost and interpretability.
To achieve this, we selected a range of extractive and abstractive summarization techniques, including heuristic-based approaches and LLMs, to evaluate their relative performance. Additionally, we fine-tuned the Mistral-7B model on financial text and assessed its performance on Financial News Summarization dataset~\cite{zhou-etal-2021-trade}. The following subsections provide a detailed description of the summarization approaches and the methodology adopted in our study.

\subsection{Approaches}
\label{subsec:approaches}

For extractive summarization, we selected the following approaches: TextRank, LexRank, Lead-$n$ baseline, DistilBERT and MatchSum.
TextRank~\cite{mihalcea-tarau-2004-textrank} is a graph-based ranking model inspired by PageRank, which selects key sentences based on their centrality in a similarity graph. LexRank~\cite{lexrank} is another graph-based method utilizing eigenvector centrality to rank sentences based on their relevance within the document. Lead-$n$ Baseline is a simple heuristic that selects the first $n$ sentences of the document as the summary, serving as a strong baseline in domains where the most relevant information tends to appear early in the text.

DistilBERT~\cite{Sanh2019DistilBERTAD} is a distilled version of the BERT model, used here for extractive summarization by leveraging contextualized vector representations of sentences. These representations are compared to that of the entire document to select the most semantically relevant sentences.
Finally, MatchSum~\cite{DBLP:journals/corr/abs-2004-08795} is a model originally designed to evaluate the quality of summaries, but adapted here for extractive use. It assigns a relevance score to each sentence in the document based on its similarity to a global representation of the text, thereby enabling the construction of a summary from the highest-scoring sentences.\\

For abstractive summarization, we evaluated several state-of-the-art LLMs in their smaller versions (7B and 8B parameters) to ensure efficiency while maintaining strong generative capabilities. The selected models include: \textrm{Mistral-7B-Instruct-v0.3},\footnote{\url{https://huggingface.co/mistralai/Mistral-7B-Instruct-v0.3}} and \textrm{Meta-Llama-3-8B-Instruct},\footnote{\url{https://huggingface.co/meta-llama/Meta-Llama-3-8B-Instruct}} \textrm{pegasus-xsum}~\cite{pmlr-v119-zhang20ae}, \textrm{t5-small}~\cite{2020t5}, \textrm{bart-large-xsum}~\cite{DBLP:journals/corr/abs-1910-13461}, \textrm{gpt-4o-mini-2024-07-18}~\cite{OpenAIChatGPT}, and \textrm{DeepSeek-R1-Distill-Qwen-7B}~\cite{deepseekai2025deepseekv3technicalreport}.
Additionally, we fine-tuned the \textrm{Mistral-7B-Instruct-v0.3} model using LoRA (Low-Rank Adaptation)~\cite{pontes-etal-2024-l3itc}, focusing on improving summarization accuracy for financial news articles.

\subsection{Dataset}
\label{subsec:corpus}

The dataset used in this study consists of financial news articles and their corresponding concise summaries \cite{xie-etal-2024-finnlp}. It includes 8,000 training instances and 2,000 test instances, providing a robust benchmark for evaluating summarization techniques in the financial domain. Each example in the dataset consists of four components: id (a unique identifier for the instance), text (the full financial news article), prompt (a predefined instruction guiding the summarization task), and summary (the ground-truth abstractive summary).


Table~\ref{tab:dataset_stats} presents key statistical characteristics of the dataset, including the average number of characters, words per document, words per sentence, and the number of sentences per document. These statistics highlight the relatively short length of summaries compared to the original documents, emphasizing the need for effective compression while preserving critical financial information. 

\begin{table}[!h]
\centering
\small  
\setlength{\tabcolsep}{4pt} 
\begin{tabular}{|l|c|c|c|c|}
\hline
\textbf{Dataset} & \textbf{\#chars} & \textbf{\#words} & \textbf{words/sent} & \textbf{\#sents} \\\hline
Train texts & 4847.99 & 710.06 & 29.84 & 23.8 \\\hline
Train sums & 143.08 & 20.71 & 19.67 & 1.05 \\\hline
Test texts & 4881.26 & 716.41 & 30.74 & 23.3 \\\hline
Test sums & 146.65 & 21.23 & 20.05 & 1.06 \\\hline      
\end{tabular}
\caption{Statistics of the Financial News Summarization dataset.} 
\label{tab:dataset_stats}
\end{table}

\vspace{-1cm}
\subsection{Summary generation}
\label{subsec:generation}

As shown in Table~\ref{tab:dataset_stats}, this dataset was chosen precisely because the summaries in the training dataset typically consist of a single sentence on average, which perfectly matches our goal of producing very concise summaries. Consequently, we configure our extractive methods to select only one sentence per summary, in order to remain consistent with this format.



For abstractive approaches, we generate one version of summaries: one-sentence summary. Therefore, we use the following prompt for LLMs: \textit{"You are given a financial text that consists of multiple sentences. You must summarize this text in only one short sentence. Text: \{text\}"}.

\textrm{FT-Mistral-7B-Instruct-v0.3} was fine-tuned on the training dataset using LoRA  with a rank of 16, allowing for efficient parameter updates while preserving the pre-trained model's knowledge. The fine-tuning process was conducted for 500 steps with a learning rate of 6e-5, optimizing the model to better capture the linguistic and structural patterns of financial news summaries.

\subsection{Metrics}
\label{subsec:metrics}

To assess the quality of generated summaries, we employ two widely used evaluation metrics: ROUGE (1, 2, 4 and L)~\cite{lin-2004-rouge} and BERTScore~\cite{zhang2020bertscoreevaluatingtextgeneration}. By combining lexical-based (ROUGE) and semantic-based (BERTScore) metrics, we aim to provide a comprehensive assessment of summarization performance, ensuring both content relevance and linguistic quality.

\section{Results}
\label{sec:results}


Table~\ref{tb:evaluation} presents the evaluation results of several summarization approaches on the test dataset. Extractive methods, notably Lead-1 and MatchSum, establish a strong baseline, outperforming unsupervised graph-based methods (TextRank, LexRank), some pretrained models such as DistilBERT, as well as most abstractive models. This effectiveness can be explained by the nature of financial articles, where key information is often located early in the text.

\color{black}

\begin{table}[!h]
\centering
\begin{tabular}{|l|c|c|c|c|c|}
\hline
\textbf{Approach/Model}              & \textbf{R-1}   & \textbf{R-2}   & \textbf{R-4}   & \textbf{R-L}   & \textbf{BERTScore} \\ \hline
\multicolumn{6}{|c|}{\textit{Extractive summaries (1 sentence summary)}} \\ \hline
Lead-1                      & 0.247 & 0.111 & 0.035 & 0.214 & 0.588       \\ \hline
MatchSum                    & 0.241 & 0.108 & 0.035 & 0.208 & 0.583     \\ \hline
LexRank                     & 0.164 & 0.057 & 0.015 & 0.133 & 0.509     \\ \hline
TextRank                    & 0.142 & 0.047 & 0.012 & 0.112 & 0.494     \\ \hline
DistilBert                  & 0.138 & 0.036 & 0.008 & 0.113 & 0.491     \\ \hline
\multicolumn{6}{|c|}{\textit{Abstractive summaries (1 sentence summary)}} \\ \hline
T5-small & 0.175  & 0.065 & 0.017 & 0.156 & 0.551     \\ \hline
Pegasus-xsum & 0.198  & 0.082 & 0.024 & 0.172 & 0.536     \\ \hline
Bart-large-xsum & 0.220  & 0.092 & 0.024 & 0.191 & 0.543     \\ \hline
DeepSeek-R1-Distill-Qwen-7B & 0.220  & 0.082 & 0.018 & 0.189 & 0.587     \\ \hline
Meta-Llama-3-8B-Instruct    & 0.206 & 0.079 & 0.020  & 0.178 & 0.572     \\ \hline
Mistral-7B-Instruct-v0.3    & 0.226 & 0.087 & 0.021 & 0.191 & 0.594     \\ \hline
gpt-4o-mini-2024-07-18      & 0.254 & 0.103 & 0.022 & 0.223 & 0.619     \\ \hline
\multicolumn{6}{|c|}{\textit{Fine-tuned LLM (1 sentence summary)}} \\ \hline
FT-Mistral-7B-Instruct-v0.3 & \textbf{0.514}  & \textbf{0.334} &\bf 0.170 & \textbf{0.480}  & \textbf{0.728}     \\ \hline
\end{tabular}
\caption{Summary evaluation on the test dataset. ROUGE and BERT scores are presented. Extractive and abstractive summarization models are compared using ROUGE (R-1, R-2, R-4, R-L) and BERTScore metrics. The best-performing results are in bold.}
\label{tb:evaluation}
\end{table}

Indeed, extractive methods avoid the factual inaccuracies frequently encountered by abstractive models, as they select sentences directly from the source text. However, they lack flexibility to rephrase or reorganize content, which can lead to redundancies or omissions of important information found later in the document.

Conversely, abstractive models produce more fluent and concise summaries by synthesizing information, but large generalist models often struggle to master the terminology and context specific to finance. As a result, several popular abstractive models (DeepSeek, Meta-Llama-3, T5-small, Pegasus-xsum, Bart-large-xsum) perform worse than Lead-1. Moreover, they can generate factual inconsistencies, limiting their reliability in contexts that require rigor and precision. To address these limitations, some approaches restrict the output to a single sentence to reduce hallucinations. For example, GPT-4o-mini outperforms Lead-1 with a BERTScore of 0.619 compared to 0.588.

Moreover, abstractive approaches can sometimes generate factual inconsistencies, making them less reliable for applications where rigor and precision are essential.
In our context, abstractive models are constrained to produce single-sentence summaries, thereby helping to reduce the risk of hallucinations. Some models thus manage to outperform extractive methods. For example, GPT-4o-mini achieves a BERTScore of 0.619, surpassing Lead-1 (0.588).

Finally, fine-tuned FT-Mistral-7B-Instruct-v0.3 outperforms all other approaches, achieving a remarkable 100 \% improvement over extractive methods in R-1 (0.514 vs. 0.247 for Lead-1) and significantly higher BERTScore (0.728). This demonstrates the effectiveness of fine-tuning in adapting LLMs to financial text, allowing them to better capture domain-specific terminology, structure, and summarization patterns. Fine-tuning not only improves accuracy but also mitigates issues related to hallucinations and irrelevant information generation.

These results highlight the trade-offs between extractive and abstractive approaches. Extractive methods remain solid baselines due to their reliability and alignment with the structure of financial news articles. However, some abstractive models offer a better balance between fluency and conciseness. Most importantly, fine-tuning a domain-specific LLM brings substantial gains, proving that generalist LLMs struggle to effectively summarize financial texts but can achieve state-of-the-art performance once adapted to the task.

The results presented in this study are based on the Financial News Summarization dataset, designed for evaluating financial news summarization. However, this dataset presents several important limitations, notably the presence of reference summaries that are difficult to use, sometimes incoherent or insufficiently informative compared to the source texts. These shortcomings, illustrated in Table~\ref{example1}, may compromise the reliability of automatic evaluations by biasing metrics such as ROUGE and BERTScore.
The scores obtained by two contrasting methods highlight this issue clearly:
%
\textbf{Lead-1} achieves a BERTScore of 0.2077, ROUGE-\{1,L\} scores of 0.054, and zero scores for ROUGE-\{2,4\}.
\textbf{FT-Mistral-7B-Instruct-v0.3} reaches a BERTScore of 1, along with perfect scores across all ROUGE metrics.
%
Paradoxically, however, the summary achieving these perfect scores is, in practice, unusable.
Beyond evaluation, these issues also have a direct impact on supervised learning. The quality of training data directly influences the quality of the resulting models. A model, even one that appears technically performant, may still be limited if it has been trained on low-quality annotations. These results should therefore be interpreted with caution: the high scores achieved by FT-Mistral 
on Financial News Summarization dataset do not necessarily guarantee strong real-world generalization. This highlights the need for better-constructed and more reliable datasets to support the development of robust automatic summarization systems in the financial domain.

\begin{table}[h]
\centering
\begin{tabular}{|l|p{10cm}|}
\hline
\bf model & \bf Text input/Summaries \\ \hline
raw\_text & \footnotesize FORM 8.3 IRISH TAKEOVER PANEL DISCLOSURE UNDER RULE 8.3 OF THE IRISH TAKEOVER PANEL ACT, 1997, TAKEOVER RULES, 2013 DEALINGS BY PERSONS WITH INTERESTS IN RELEVANT SECURITIES REPRESENTING 1\% OR MORE 1. KEY INFORMATION Name of person dealing (Note 1) Farallon Capital Management, L.L.C. on behalf of funds managed jointly with Farallon Partners, L.L.C Company dealt in Aon PLC Class of relevant security to which the dealings being disclosed relate (Note 2) US~\$0.01 ordinary shares Date of dealing 12 April 2021 2. INTERESTS AND SHORT POSITIONS (a)Interests and short positions (following dealing) in the class of relevant security dealt in (Note 3) Long Short Number (\%) [...] 
\\ \hline \hline
gold & \footnotesize Form 8.3 - Aon PLC \\ \hline
lead\_1 & \footnotesize FORM 8.3 IRISH TAKEOVER PANEL DISCLOSURE UNDER RULE 8.3 OF THE IRISH TAKEOVER PANEL ACT, 1997, TAKEOVER RULES, 2013 DEALINGS BY PERSONS WITH INTERESTS IN RELEVANT SECURITIES REPRESENTING 1\% OR MORE 1. \\ \hline
mistral\_1 & \footnotesize Farallon Capital Management purchased 24,000 US~\$0.01 ordinary shares of Aon PLC on April 12, 2021, increasing their long position to 2,468,814 shares. \\ \hline
chatgpt\_1 & \footnotesize Farallon Capital Management disclosed its dealings in Aon PLC, reporting long and short positions in relevant securities as of April 12, 2021. \\ \hline
mistral\_ft & \footnotesize Form 8.3 - Aon PLC \\ \hline
\end{tabular}
\caption{\label{example1}Summaries generated by several approaches for a regulatory financial article.}
\end{table}

\section{Conclusion}
\label{sec:CONC}

Abstractive approaches demonstrated strong capabilities in capturing the main facts from both simple and complex source articles when generating summaries, shown in Table~\ref{example2} and~\ref{example3}. However, their high computational cost and slow processing time make them impractical for summarizing large volumes of articles exclusively using these methods. In contrast, extractive approaches performed well in identifying key information from short, well-structured financial articles but showed significant limitations when applied to more complex texts, such as financial reports. This highlights the inherent constraints of extractive methods. Additionally, fine-tuned large language models (LLMs) generated title-based summaries aligned with the Financial News Summarization dataset. While these summaries achieved the highest ROUGE and BERTScore metrics, they occasionally omitted crucial information and were less readable. In conclusion, extractive methods are effective for summarizing short, well-structured articles, while LLM-based abstractive models are better suited for handling complex financial documents.

Future work will focus on extending the evaluation to datasets such as earnings calls and 10-K filings, enabling more robust comparisons across approaches that must handle financial documents of varying length and complexity. To better process this diversity, we will enhance extractive baselines with named entity recognition and event extraction to capture key facts more effectively. In parallel, we will investigate domain-adapted LLMs, incorporating larger training corpora and retrieval-augmented techniques to improve factuality and coherence in summarizing long and complex financial documents. Finally, employing LLMs as judges offers a scalable alternative for quality evaluation, covering dimensions such as informativeness, readability, and factual coverage, when annotated data are scarce or costly to obtain~\cite{chiang-lee-2023-large}.

\section{Annex: Output examples}

{\tiny
\begin{table}[h]
\centering
\begin{tabular}{|l|p{10.5cm}|}
\hline
\bf model & \bf Text input/Summaries \\\hline
raw\_text & \footnotesize Technavio has been monitoring the all-season tire market in Europe and it is poised to grow by USD 3.42 billion during 2020-2024, progressing at a CAGR of almost 9\% during the forecast period. The report offers an up-to-date analysis regarding the current market scenario, latest trends and drivers, and the overall market environment [...]
\\\hline \hline
gold & \footnotesize All-season Tire Market in Europe to Reach USD 3.42 Billion by 2024, Bridgestone Corp. and Continental AG Emerge as Key Contributors to Growth | Technavio\\\hline
lead\_1 & \footnotesize Technavio has been monitoring the all-season tire market in Europe and it is poised to grow by USD 3.42 billion during 2020-2024, progressing at a CAGR of almost 9\% during the forecast period. \\\hline
mistral\_1 & \footnotesize The all-season tire market in Europe is expected to grow by USD 3.42 billion from 2020-2024, at a CAGR of nearly 9\%, despite the COVID-19 pandemic \\\hline
chatgpt\_1 & \footnotesize The all-season tire market in Europe is expected to grow by USD 3.42 billion from 2020 to 2024, driven by various market trends and the impact of COVID-19. \\\hline
mistral\_ft & \footnotesize All-Season Tire Market in Europe 2020-2024 | Growing Adoption of Nanotechnology to Boost the Market Growth | Technavio
\\\hline
\end{tabular}
\caption{\label{example2}Summaries generated by several approaches for a press release article related to market research.}
\end{table}
}

{\tiny
\begin{table}[h]
\centering
\begin{tabular}{|l|p{10.5cm}|}
\hline
\bf model & \bf Text input/Summaries \\\hline
raw\_text & \footnotesize The V Foundation for Cancer Research, a top-rated cancer research charity, announces \$10.3 million in grant funding for 2020. It was a year unlike any other, and that was reflected in the composition of the 42 grants awarded. Of significance, three Abeloff V Scholars tied for the highest grant score, and the Foundation expanded the V Clinical Scholar program, after a successful pilot program was initiated at MD Anderson Cancer Center. Total grant funding by the V Foundation has now surpassed \$260 million [...] 
\\\hline \hline
gold & \footnotesize \$10.3 Million Awarded by the V Foundation in 2020 All-Time Grant Total More Than \$260 Million \\\hline
lead\_1 & \footnotesize The V Foundation for Cancer Research, a top-rated cancer research charity, announces \$10.3 million in grant funding for 2020. \\\hline
mistral\_1 & \footnotesize The V Foundation for Cancer Research awarded \$10.3 million in grants for 2020, funding innovative research projects and supporting a new generation of cancer researchers, with a focus on tackling challenging cancer problems, promoting equity in treatment, and advancing cancer treatment and cures. \\\hline
chatgpt\_1 & \footnotesize The V Foundation for Cancer Research announced \$10.3 million in grants for 2020, surpassing \$260 million in total funding to support innovative cancer research and promote equity in treatment. \\\hline
mistral\_ft & \footnotesize V Foundation Awards \$10.3 Million in Cancer Research Grants
\\\hline
\end{tabular}
\caption{\label{example3}Summaries generated by several approaches for a grant funding press release.}
\end{table}
}

\color{black}

\section{Credits}
This study was funded by Avignon Université, France (grant Agorantic Project CONTROV).

\bibliographystyle{apalike}
\bibliography{references.bib}

\end{document}